# Precursor phase with full phonon softening above the charge-density-wave phase transition in *2H*-TaSe$_2$

Xingchen Shen[1,2], Rolf Heid[1], Roland Hott[1], Björn Salzmann[3], Marli dos Reis Cantarino[3,4], Claude Monney[3], Ayman H. Said[5], Bridget Murphy[6,7], Kai Rossnagel[6,7,8], Stephan Rosenkranz[9], Frank Weber[1,✉]

[1] Institute for Quantum Materials and Technologies, Karlsruhe Institute of Technology, 76021 Karlsruhe, Germany
[2] College of Physics, Chongqing University, Chongqing 401331, P. R. China
[3] Département de Physique and Fribourg Center for Nanomaterials, Université de Fribourg, 1700 Fribourg, Switzerland
[4] Instituto de Física, Universidade de São Paulo, São Paulo-São Paulo 05508-090, Brazil
[5] Advanced Photon Source, Argonne National Laboratory, Lemont, Illinois 60439, USA
[6] Institute of Experimental and Applied Physic and KiNSIS, Kiel University, 24098 Kiel, Germany
[7] Ruprecht Haensel Laboratory, Kiel University, 24098 Kiel, Germany
[8] Ruprecht Haensel Laboratory, Deutsches Elektronen-Synchrotron DESY, 22607 Hamburg, Germany Germany
[9] Materials Science Division, Argonne National Laboratory, Lemont, Illinois 60439, USA

✉ *frank.weber@kit.edu*

Research on charge-density-wave (CDW) ordered transition-metal dichalcogenides continues to unravel new states of quantum matter correlated to the intertwined lattice and electronic degrees of freedom. Here, we report an inelastic x-ray scattering investigation of the lattice dynamics of the canonical CDW compound *2H*-TaSe$_2$ complemented by angle-resolved photoemission spectroscopy. Our results rule out the *central-peak* scenario for the CDW transition in *2H*-TaSe$_2$ and provide evidence for a novel precursor phase above the CDW transition temperature $T_{CDW}$. The phase at temperatures between T* (= 128.7 K) and $T_{CDW}$ (= 121.3 K) is characterized by a fully softened phonon mode and medium-range ordered ($\xi_{corr}$ = 100 Å - 200 Å) static CDW domains. Only $T_{CDW}$ is detectable in our photoemission experiments. Thus, *2H*-TaSe$_2$ exhibits structural before electronic static order and emphasizes the important lattice contribution to CDW transitions.

## Introduction

Layered transition-metal dichalcogenides (TMD) continue to provide a rich playground for emergent physics including excitonic insulators[1-3], dimensionality-dependent correlated electronic phases[4,5] as well as intriguing magnetic properties[6] with potential applications in spintronic devices[7]. Charge-density-wave (CDW) order, a periodic modulation of the charge carrier density, is widespread in metallic TMDs[8,9]. In the seminal model by Peierls[10] an electronic instability in the presence of finite electron-phonon coupling (EPC) stabilizes the CDW ground state. While this scenario applies to several known quasi-1D materials[11-14], recent work for layered TMDs[15-18] showed that the full momentum dependence of both the electronic band structure and the EPC have to be taken into account to explain CDW formation and the existence of closely related superconducting phases[19-24].

$2H$-TaSe$_2$ is a prototypical CDW compound featuring a large periodic lattice distortion[25] and a momentum-dependent energy gap in the electronic band structure in its low-temperature state[26,27]. It is a layered material [see inset in Fig. 1] for which CDW order with a transition temperature $T_{CDW}$ = 122.3 K was reported in the 1970s [28]. On cooling through T$_{CDW}$, $2H$-TaSe$_2$ first enters a CDW phase with an incommensurate ordering wave vector $\boldsymbol{q}_{CDW}$ = (0.323,0,0),[1] which gradually evolves on cooling and reaches the commensurate value $\boldsymbol{q}_{CDW}$ = (1/3,0,0) at $T_{CDW-C}$ ≈ 90 K.[29]

As a continuous structural phase transition, a CDW transition is typically associated with a phonon mode softening to zero energy at $\boldsymbol{q}_{CDW}$ on cooling towards T$_{CDW}$. Thus, the investigation of this soft phonon mode can yield important insights about the underlying mechanism governing the phase transition and has been indispensable to understand the physics in many CDW materials[14,15,17,18]. In $2H$-TaSe$_2$, the CDW soft phonon mode corresponds to the longitudinal acoustic (LA) phonon propagating along the [100] direction and has been investigated by inelastic neutron scattering (INS) experiments in the 1970s [25,28]. Surprisingly, the observed phonon softening was not complete, i.e., the energy of the LA mode at $\boldsymbol{q}_{CDW}$ softened from above 7 meV at room temperature only to 4.5 meV at $T_{CDW}$. However, a resolution-limited, static central peak at zero energy transfer developed already below 150 K, well above $T_{CDW}$. Similar observations had been made earlier in ferroelectric SrTiO$_3$ [30-32] and more recently in quasi-1D CDW compounds[12] as well as in cuprates[33]. The origin of the central peak is not fully understood[32] though it is often ascribed to order fluctuations.[12,33] Currently, $2H$-TaSe$_2$ is considered as one of the earliest materials for which a central-peak was observed, although the authors of the original INS study[25] pointed out that the coarse momentum resolution in their experiments may have obscured a better view of the lattice dynamics. Hence, the origin of the central-peak in $2H$-TaSe$_2$ remains unclear but is highly relevant, e.g., for our understanding of other materials featuring CDW order such as quasi-1D (TaSe$_4$)$_2$I,[34] NbSe$_3$,[35] ZrTe$_3$,[12] and underdoped cuprates[33,36].

Here, we employ inelastic x-ray scattering (IXS) with *meV* resolution to study the CDW soft phonon mode in $2H$-TaSe$_2$. The much higher momentum resolution of IXS compared to INS enables the detailed study of the soft-mode properties as function of wave vector as well as temperature. Combined with the energy resolution of 1.5 meV, IXS enables an unambiguous separation between the static superlattice peak (resolution limited in energy) and in- or quasi-elastic (not resolution limited) scattering from the soft phonon mode. We observe a full phonon softening of the LA phonon mode at $\boldsymbol{q}_{CDW}$ = (0.323,0,0) and that the static CDW superlattice peak intensity rises strongly only on cooling to temperatures below that of the full phonon softening. This is in contrast to previous results[25] and rules out the *central-peak* scenario for $2H$-TaSe$_2$. Yet, the most surprising finding is that the phonon softening occurs at a temperature T* = 128.7 K, i.e., well above T$_{CDW}$ = 121.3 K. Analysis of high momentum resolution scans at E = 0 reveals that the phase at T$_{CDW}$ ≤ T ≤ T* is characterized by lattice fluctuations, observed as the overdamped soft phonon mode, coexisting with static but only medium-range CDW order.

## Methods

The IXS experiments were carried out at the 30-ID beamline, HERIX spectrometer[37], at the Advanced Photon Source, Argonne National Laboratory. Phonon excitations measured in constant-momentum scans were approximated by damped harmonic oscillator (DHO) functions[38] convoluted with a pseudo-voigt resolution function (full-width at half-maximum (FWHM) = 1.5 meV). The resolution function was further used to approximate resolution limited elastic scattering at zero energy transfer. Measurements were done at scattering wave vectors $\mathbf{Q} = \boldsymbol{\tau} - \mathbf{q}$, where $\boldsymbol{\tau}$ is a reciprocal lattice vector and $\mathbf{q}$ is the reduced wave vector in the 1st Brillouin zone. Wave vectors are expressed in reciprocal lattice units (r.l.u.) $(2\pi/a, 2\pi/b, 2\pi/c)$ with the lattice constants $a = b = 3.44$ Å and $c = 12.7$ Å of the high-temperature hexagonal unit cell (#194). All measurements were done in the Brillouin zone adjacent to $\boldsymbol{\tau} = (3,0,1)$. In the following, results are presented in reduced wave vectors $\mathbf{q} = \boldsymbol{\tau} - \mathbf{Q}$. Measurements were done at constant energy or constant momentum transfer. For the former, we employed the best momentum resolution possible on the HERIX spectrometer ($\Delta q$ = 0.018 Å$^{-1}$) by decreasing the effective size of the backscattering analyzers to a circular diameter of 18 mm (FWHM) compared to 95 mm opening in energy scans at constant momentum transfer ($\Delta q$ = 0.09 Å$^{-1}$). For more details on the HERIX setup and data analysis see the Supplemental Material (SM). Complementary ARPES measurements were performed at the Bloch endstation of the R1 synchrotron at the MAXIV institute in Lund, Sweden. More experimental details and results are given in SM.

---

[1] All wave vectors are given in reciprocal lattice units (see Methods).

**Results**

The strategy of our IXS experiment is outlined in Figure 1: The LA mode at $\mathbf{q}$ = (h,0,0) shows a pronounced softening close to the temperature of the phase transition at $\mathbf{q}_{CDW}$ [black lines in Fig. 1(a)] where eventually the CDW superlattice peak (red dot at zero energy transfer) emerges. We performed energy scans at constant momentum transfer [vertical (green) arrows] and momentum scans at zero energy transfer [horizontal (blue) arrow]. Typical results for three different scenarios [vertical and horizontal arrows in Fig. 1(a)] are sketched in Figures 1(b)-(d). At $\mathbf{q} \neq \mathbf{q}_{CDW}$, the LA mode can already have a low energy, e.g., $E_{phon} \approx 4$ meV. Taking into account the energy resolution and damping due to electron-phonon coupling, the resulting scan shows broad phonons peaked at $\pm E_{phon}$ and with finite scattering intensity even at zero energy transfer [Fig. 1(b)]. Constant momentum scans at $\mathbf{q} = \mathbf{q}_{CDW}$ and T ≈ T$_{CDW}$ typically show a resolution-limited signal corresponding to the rising static CDW superlattice peak [solid (red) line in Fig. 1(c)] easily distinguishable from the broad scattering from the damped soft phonon mode [dashed (blue) line]. Due to the small phonon scattering intensities, constant momentum scans employ a relatively broad momentum resolution, $\Delta q$ = 0.09 Å$^{-1}$, not suited to investigate the correlation length of the rising CDW superlattice peak. Therefore, we performed additional momentum scans at zero energy transfer across $\mathbf{q} = \mathbf{q}_{CDW}$ with high resolution, $\Delta q$ = 0.018 Å$^{-1}$. Generally, it is not clear which part of the scattering intensity at zero energy transfer originates from the static central peak and which part is due to the (nearly) completely damped LA phonon [Fig. 1(d)]. Yet, the combined analysis of both types of scans allows the unambiguous determination of the temperature dependence of the elastic CDW superlattice peak concomitant with that of the soft phonon mode.

Figure 2 illustrates the dispersion of LA mode [Figs. 2(a)-(c) and 2(d)-(f)] and highlights its strong temperature dependence at $\mathbf{q}$ = (0.325,0,0) [≈ $\mathbf{q}_{CDW}$ = (0.323,0,0)] [Figs. 2(b)(e)]. The energy of the LA mode [Fig. 3(a)] shows a clear soft-mode behavior [solid red line in Fig. 3(a)] as a function of temperature. Surprisingly, the softening is complete already just below 130 K, well above the reported CDW transition temperature[25] of about 122 K [vertical blue dashed line in Fig. 3(a)]. Data obtained at a slightly offset wave vector [open circles in Fig. 3(a)] allow measurements above and below the CDW transition and clearly reveal the hardening of the LA mode on cooling into the CDW phase. Fitting a power law of the form |T-T*|$^\delta$ [red solid line in Fig. 3(a)] to the temperature dependence of the soft mode for T ≥ 130 K yields δ = 0.32(2) and T* = 128.7(3) K [vertical red dashed line in Fig. 3(a)]. Below T*, the energy of the soft mode stays close to zero. The observed value of δ is lower than the mean-field one, $\delta_{MF}$ = 0.5, observed e.g. for the CDW transition in the iso-structural 2H-NbSe$_2$.[15] Since mean-field theory neglects fluctuations of the order parameter, δ = 0.32(2) indicates the presence of strong order parameter fluctuations in 2H-TaSe$_2$. We expect reduced fluctuations far away from the transition and, indeed, the soft-mode energies for T ≥ 175 K follow mean-field behavior, which means that the square of the phonon energies depend linearly on temperature [see Fig. 3(b)]. A linear fit for T ≥ 175 K extrapolates zero energy for the soft mode at T ≈ 85 K, which is close to T$_{C-IC}$ ≈ 88 K,[28,29] the temperature at which 2H-TaSe$_2$ acquires a commensurate CDW.

Before we discuss the temperature range T ≈ T* in more detail, we present the dispersion and the damping of the LA soft mode in Figures 3(c) and (d), respectively. We find a V-shaped dispersion and a sharply peaked damping centered at $\mathbf{q}_{CDW}$ at T = 130 K [black spheres in Fig. 3(c)] again in contrast to 2H-NbSe$_2$, where the LA soft mode displays a much broader, U-shaped dispersion at the respective $T_{CDW}$.[15,39] Phonon energies of the LA mode rise much faster going away from $q_{CDW}$ in *2H*-TaSe$_2$ than in *2H*-NbSe$_2$. A classic V-shaped dispersion [Fig. 3(c)] is expected when the EPC of the soft mode is dominated by a Fermi surface (FS) nesting. In that case, the strong increase of electronic decay channels at the nesting wave vector leads to sharply localized phonon anomalies. The dispersion of *2H*-TaSe$_2$ is sharper than in *2H*-NbSe$_2$, in which FS nesting was shown to be completely absent[40], but not as sharp as in some quasi-1D compounds, such as KCP[11] and ZrTe$_3$.[12] This is in line with results from angle-resolved photoemission spectroscopy (ARPES) revealing a moderately strong FS nesting in *2H*-TaSe$_2$[41]. The range of wave vectors over which the LA mode shows significant renormalization is about 0.08 – 0.1 r.l.u. along the [100] direction, in reasonable agreement with the coherence length of 14 Å estimated from specific heat measurements.[42]

Figure 4 shows energy scans at $q_{CDW}$ taken for temperatures $T_{CDW} \leq T \leq T^*$ along with an analysis of the temperature dependence of the observed static CDW superlattice peak. The results show that the CDW superlattice peak [dash-dotted red lines in Figs. 4(a)-(c)] increases in this temperature range by another factor of 10. Yet, the spectrum at T = 122.5 K still shows similar spectral weights of the CDW superlattice peak (dash-dotted red line) and the soft phonon mode [solid blue line in Fig. 4(c)]. Cooling further by only ΔT = 1 K, the scattering at zero energy jumps by a factor of 20 [inset of Fig. 4(c)] and we will see further below that this jump marks the onset of long-range CDW order. However, the strong increase of the superlattice peak prohibits further analysis of in- or quasi-elastic scattering at lower temperatures with the given energy resolution, because the much weaker quasi-elastic scattering is hidden under the tail of the elastic peak. The temperature dependences of the soft mode's energy, its intensity and that of the CDW superlattice peak are summarized in Figures 4(d) and (e), respectively. The intensity of the soft phonon mode dominates the spectra except for close to and, of course, below $T_{CDW}$. Nevertheless, our results clearly show that a small static CDW superlattice peak is present already at T ≈ 130 K. The increase of the CDW superlattice peak intensity accelerates sharply below $T^*$ [vertical red dashed line in Fig. 4(e), *note the break of the vertical scale*] before it jumps by another factor of 20 crossing $T_{CDW}$ [see inset in Fig. 4(c)].

Our results identify a precursor phase at $T_{CDW} \leq T \leq T^*$ characterized by the simultaneous presence of soft-mode lattice fluctuations and static CDW order. A critically damped phonon indicates that there is quasi-zero energy needed for small lattice distortions which are symmetrically related to the phonon oscillation pattern. On the other hand, the intensity of a superlattice peak is proportional to the square of the order parameter, i.e., the atomic displacement from the high-temperature equilibrium position. Thus, the presence of a weak CDW superlattice peak indicates small but finite displacements. A possible interpretation is that there exist small domains of static CDW order with a finite correlation length $\xi_{corr}$, in a matrix of material with critical lattice fluctuations. On cooling, one would expect that $\xi_{corr}$ increases and finally diverges at the temperature at which the sample reaches long-range CDW order, i.e., at $T_{CDW}$.

However, the momentum resolution employed for energy scans ($\Delta q$ = 0.09 Å$^{-1}$) is too coarse for a detailed analysis of the CDW correlation length $\xi_{corr}$. Therefore, we performed momentum scans at constant energy transfer of E = 0 [indicated by the horizontal blue arrow in Fig. 1(a) and illustrated in Fig. 1(d)] with an improved momentum resolution of $\Delta q$ = 0.018 Å$^{-1}$ (see methods). Raw data at selected temperatures are shown in Figure 5(a). Data at T = 122 K [red spheres in Fig. 5(a)] show the rise of scattering intensity for a broad momentum range $q$ = (0.3 – 0.35,0,0) on cooling. Just one degree below, at T = 121 K, we find a more than twenty-fold increase of the scattering intensity centered at $q_{CDW}$ = (0.323,0,0). On further cooling the intensity increases further and the peak position moves towards the commensurate wave vector $q_{CDW,C}$ = (1/3,0,0) illustrated by the data for T = 80 K [orange spheres in Fig. 5(a)]. From a peak fit we obtain the detailed temperature dependences of the integrated intensity [Fig. 5(b)], the peak position $q_{CDW}$ [Fig. 5(c)] and its line width $\Gamma_{exp,FWHM}$ [Fig. 5(d)]. The integrated peak intensity follows a power law of the form $|T-T_{CDW}|^{2\beta}$ for T ≤ 121 K with $T_{CDW}$ = 121.3(2) K and β = 0.21(1) [red line in inset of Fig. 5(b)]. At $T < T_{CDW}$, the intensity continues to increase and $q_{CDW}$ shifts to the commensurate value $q_{CDW,C}$ = (1/3,0,0) at temperatures close to $T_{C-IC}$ ≈ 88 K[29] [inset in Fig. 5(c)]. $\Gamma_{exp,FWHM}(T)$ shows that peaks are resolution limited for data taken at T ≤ 121 K. A power law fit of the form $(T-T_{CDW})^\delta$ to $\Gamma_{exp,FWHM}(T)$ for T ≥ 122 K corroborates the transition temperature to long-range CDW order $T_{CDW}$ = 121.3(2) K in good agreement with the literature.[25,28,29,42] A more detailed analysis which takes into account the experimental resolution and estimates the phononic background (see Fig. S4 and SM) shows that the correlation length $\xi_{corr}$ of the static CDW domains in the precursor phase increases along the [100] on cooling from 100 Å to 200 Å just above $T_{CDW}$. Values along [010] are essentially the same while the correlation length along [001] is 4-5 times smaller. Thus, the precursor phase in *2H*-TaSe$_2$ is characterized by medium-range-sized CDW domains, which only form a long-range CDW ordered state at T ≤ $T_{CDW}$.

The precursor phase has not been observed previously by other techniques. INS experiments[25,28] did not have sufficient momentum resolution while XRD[29] was energy integrated and so could not distinguish between static CDW and soft-mode intensity contributions. We performed synchrotron-based ARPES to cross-check results on our own samples with previous reports[27]. Measurements were done with an incident photon energy of 80 eV (resolution < 10 meV) over a temperature range 132 K ≥ T ≥ 113 K, i.e., cooling across T* and $T_{CDW}$. Technical details and detailed results of our analysis are given in SM. Here, we summarize, that - in agreement with previous reports – the electronic band structure exhibits a pseudo-gap (as defined in Ref. [27], see SM and Fig. 7 for more details), which increases slowly on cooling in the upper part of the temperature range [open squares in Fig. 4(e)]. The increase of the pseudo-gap on cooling accelerates by a factor 4 at lower temperatures and linear fits within the two temperature regions [dashed lines in Fig. 4(e)] intersect very close to $T_{CDW}$ = 121.3 K. However, no particular response is detectable on crossing T*. The presence of the pseudo-gap in *2H*-TaSe$_2$ has been reported up to nearly room temperature[41] as well as in the high-temperature phase of *2H*-NbSe$_2$.[43] The latter report found that the pseudo-gap phase in *2H*-NbSe$_2$ at T > $T_{CDW}$ is related to incoherent CDW fluctuations. Long-range CDW order only sets in once phase coherence is established at T < $T_{CDW}$. On the other hand, electrons adjust quasi-instantaneously to lattice motions. It is argued for the case of *2H*-NbSe$_2$[43] that fluctuations always yield a pseudo-gap irrespective of the time-scale of the fluctuations. Therefore, we argue that electronic degrees of freedom do not distinguish between lattice fluctuations with a finite life time, i.e., the overdamped soft mode, and static correlations but follow the evolution of the energy-integrated signal as probed by standard x-ray diffraction[29]. Consequently, the pseudo-gap is rather insensitive to T* but strongly responds to the large increase of scattering intensity on cooling below $T_{CDW}$.

**Discussion**

The presented IXS investigation puts the CDW transition in *2H*-TaSe$_2$ in a new light in that there is a full phonon softening. As already suggested in the original publications[25,28], the coarse momentum resolution of INS in concert with the V-shaped soft mode dispersion prevented the observation of the full softening. Moreover, our analysis shows that the tail of the scattering at T ≥ $T_{CDW}$ [Fig. 4(b)], also observed in synchrotron XRD[29] is dominated by scattering from the soft LA mode for T ≥ T* [see Fig. 4(e)]. The central peak at T > T* is a factor of twenty smaller than the soft mode intensity [see Fig. 2(a)]. For *2H*-TaSe$_2$, we can therefore rule out the central-peak scenario as it was observed, *e.g.*, in SrTiO$_3$,[30-32] and other CDW compounds.[12,33-36] One can ask the question whether other central-peak compounds might show similar surprises as *2H*-TaSe$_2$. In general, we believe that applying IXS to study the structural instabilities in materials such as SrTiO$_3$, KMnF$_3$ or Nb$_3$Sn, which have been investigated at a time when only INS was available, will provide new insights because of the good momentum resolution as well as the ability to investigate very small samples having less defects/vacancies. Reports of central-peaks in ZrTe$_3$ and other quasi-1D CDW materials are already based on IXS. Quasi-1D compounds show inherently stronger fluctuations and this could be the reason that a central peak is present while it is not in *2H*-TaSe$_2$.

*2H*-TaSe$_2$ seems to be one of the few examples among layered TMDs where FS nesting defines the periodicity of the charge modulation. For comparison, FS nesting is completely absent in iso-structural *2H*-NbSe$_2$ ($T_{CDW,NbSe2}$ = 33 K) [40] and its impact in *1T*-VSe$_2$ ($T_{CDW,VSe2}$ = 110 K) [44] was recently questioned by a study combining IXS and anharmonic *ab-initio* calculations[18] emphasizing the 3D character of the CDW transition in *1T*-VSe$_2$. Another difference is that the phonon softening in both *2H*-NbSe$_2$ [15] and *1T*-VSe$_2$ [18] feature critical exponents of δ ≈ 0.5, i.e., the mean-field value. In a detailed study on 2H-NbSe2, some of us did not see any indication of a central peak above $T_{CDW}$.[15] δ = 0.32 in *2H*-TaSe$_2$ [see Fig. 3(d)] indicates the presence of fluctuations. Both, the decisive role of FS nesting and the presence of fluctuations are in line with previous studies in *2H*-TaSe$_2$ using ARPES[41] and high-momentum resolution x-ray diffraction[29], which assigned temperature dependent nesting features to competing CDW fluctuations reflected in the gradual evolution of ***q***$_{CDW}$ on cooling from an incommensurate to a commensurate value for $T \leq T_{CDW}$ [see Fig. S1(b)]. Similarly, our analysis of the phonon softening indicates deviations from mean-field behavior for T ≤ 150 K and estimates the full phonon softening in absence of fluctuations close to $T_{C-IC}$ = 88 K [see inset in Fig. 2(c)].

However, the most striking feature of the CDW transition in *2H*-TaSe$_2$ is the precursor phase for $T_{CDW}$ = 121.3 K ≤ T ≤ T* = 128.7 K characterized by a fully damped soft phonon mode and a gradually rising static CDW superlattice peak (see Fig. 3). Though the central-peak scenario does not apply, *2H*-TaSe$_2$ acquires (at least partially) a distorted lattice without concomitant electronic signatures evidenced by previous[26,27] and our own ARPES data (see SM). Furthermore, structural probes including our own data reveal the strongest phase-transition-like signatures on cooling below $T_{CDW}$ (see Fig. 4) while changes in the electronic band structure are much more pronounced at the incommensurate-to-commensurate transition at lower temperatures[27,41]. A dominant lattice, i.e., phonon contribution to the phase transition entropy on entering the incommensurate CDW phase at $T_{CDW}$ was also suggested based on specific heat measurements[42] and theoretical considerations.[45] Thus, structural order accompanied by a (weak) pseudogap (see SM and [27]) precedes full electronic order (featuring a well-defined gap on the FS) in 2H-TaSe$_2$ in the presence of competing CDW fluctuations.

Recently, CDW order in quasi-1D ZrTe$_3$ has been shown to feature a structural transition at the typically reported $T_{CDW}$ = 63 K while electronic long-range order only sets in at $T_{LO}$ = 56 K [46]. The soft phonon mode shows a temperature dependence best described by a power law with δ ≈ 1/8 assigned to fluctuations[12] in the presence of FS nesting[13]. Yet, lattice dynamics in ZrTe$_3$ are qualitatively different to those in *2H*-TaSe$_2$ in several aspects: the softening is complete only at $T_{CDW}$, a central elastic peak rises already above $T_{CDW}$[12] and CDW order remains always incommensurate[46].

*2H*-TaSe$_2$ seems to be between mean-field-like *2H*-NbSe$_2$ and quasi-1D ZrTe$_3$ in terms of strength of FS nesting, critical exponent δ of the phonon softening and impact of fluctuations. The CDW transition in *2H*-NbSe$_2$ is a show case for momentum-dependent EPC[15-17] and it has been argued that the latter must be considered for a quantitative understanding of the CDW transition also in ZrTe$_3$ as well.[47] As *2H*-TaSe$_2$ seems to be between these two examples, the study of equally important electronic and lattice degrees of freedom presents an interesting topic for future research. Finally, the full phonon softening and medium-range CDW order above $T_{CDW}$, which, to our knowledge, is unique to 2H-TaSe$_2$ among CDW compounds, is likely a manifestation of intense order-parameter fluctuations, which are already duped responsible for the strongly temperature dependent CDW properties at $T_{CDW,IC} \leq T \leq T_{CDW}$. Indeed, the analysis of the phonon softening shows deviations from mean field behavior and, thus, indicates the pronounced impact of fluctuations in the temperature range 90 K ≤ T ≤ 150 K [see inset in Fig. 2(c)]. We expect that momentum dependent EPC and the FS topology have to be taken into account for a full understanding of CDW order in *2H*-TaSe$_2$ which is beyond the scope of the present work.

**Acknowledgments**
X. S. was supported by the Helmholtz-OCPC Postdoc Program. S.R. was supported by the Materials Sciences and Engineering Division, Office of Basic Energy Sciences, U.S. Department of Energy. B.S., M.d.R.C. and C.M. acknowledge financial support from the Swiss National Science Foundation (SNSF) Grant No. P00P2 170597. M.d.R.C. was supported by grant #2020/13701-0, São Paulo Research Foundation FAPESP. This research used resources of the Advanced Photon Source, a U.S. Department of Energy (DOE) Office of Science User Facility operated for the DOE Office of Science by Argonne National Laboratory under Contract No. DE-AC02-06CH11357. The authors gratefully acknowledge MAX IV Laboratory for time on Beamline Bloch under Proposal 20200293, as well as the support from C. Polley and G. Carbone.

**Appendix A : Inelastic x-ray scattering**

The IXS experiments were carried out at the 30-ID beamline, HERIX spectrometer[37], at the Advanced Photon Source, Argonne National Laboratory, with a focused beam size of 15 μm × 32 μm. The incident energy was 23.78 keV [48] and the horizontally scattered beam was analyzed by a set of spherically curved silicon analyzers (Reflection 12 12 12) [49]. The full width at half maximum (FWHM) of the energy and wave vector space resolution was about 1.5 meV and 0.09 Å$^{-1}$, respectively, where the former is experimentally determined by scanning the elastic line of a piece of Plexiglas and the latter is calculated from the experiment geometry and incident energy. Phonon excitations measured in constant-momentum scans were approximated by damped harmonic oscillator (DHO) functions[38] convoluted with a pseudo-voigt resolution function (FWHM = 1.5 meV). The resolution function was further used to approximate resolution limited elastic scattering at zero energy transfer. Measurements were done at scattering wave vectors $\mathbf{Q} = \boldsymbol{\tau} - \mathbf{q}$, where $\boldsymbol{\tau}$ is a reciprocal lattice vector and $\mathbf{q}$ is the reduced wave vector in the 1$^{st}$ Brillouin zone. Wave vectors are expressed in reciprocal lattice units (r.l.u.) $(2\pi/a, 2\pi/b, 2\pi/c)$ with the lattice constants $a = b = 3.44$ Å and $c = 12.7$ Å of the high-temperature hexagonal unit cell (#194). All measurements were done in the Brillouin zone adjacent to $\boldsymbol{\tau} = (3,0,1)$. Results are presented in reduced wave vectors $\mathbf{q} = \boldsymbol{\tau} - \mathbf{Q}$. We used a high-quality single crystal sample grown at the University of Kiel weighing about 5 mg (2 x 2 x 0.02 mm³). The sample was mounted in a closed-cycle refrigerator and measurements reported here were done at various temperatures $60\,\text{K} \leq T \leq 300\,\text{K}$.

Measured energy spectra at constant momentum transfer were approximated using a pseudo-Voigt function for the resolution limited elastic peak and a damped harmonic oscillator (DHO) function [38] for the phonon peaks. The DHO function was convoluted with the fit of the experimental resolution function. The DHO function is

$$S(Q,\omega) = \frac{[n(\omega)+1]Z(Q)4\omega\Gamma/\pi}{[\omega^2 - \widetilde{\omega}_q^2]^2 + 4\omega^2\Gamma^2} \quad (1)$$

where $\mathbf{Q}$ and $\omega$ are the wavevector and energy transfer, respectively, n(ω) is the Bose function, Γ is the imaginary part of the phonon self-energy, $\widetilde{\omega}_q$ is the phonon energy renormalized by the real part of the phonon self-energy and $Z(\mathbf{Q})$ is the phonon structure factor. This function covers the energy loss and energy gain scattering by a single line shape. The intensity ratio of the phonon peaks at $E = \pm\omega_q$ is fixed by the principle of detailed balance. The energy $\omega_q$ of the damped phonons is obtained from the fit parameters of the DHO function by $\omega_q = \sqrt{\widetilde{\omega}_q^2 - \Gamma^2}$ [50].

We note that IXS spectra generally contain a small, resolution-limited peak at zero energy transfer due to crystal imperfections. In our sample of *2H*-TaSe$_2$ this defect scattering peak is tiny with amplitudes < 1 count/mon×10$^5$ observed at temperatures far above T$_{CDW}$ [Figs. 2(a)-(c)], confirming its high quality. Below T < 150 K , the zero energy transfer intensity at $\mathbf{q}_{CDW}$ starts to slowly rise due to the onset of the static CDW superlattice peak. For T = 130 K, the data exhibit a resolution-limited peak at E = 0 [dash-dotted (black) line in Fig. 2(e)] with an amplitude that is ten times the value of the elastic line due to crystal imperfection observed at temperatures T > 150K. Yet, this intensity is still small compared to that of the soft phonon mode [solid (blue) line in Fig. 2(e)]. The static superlattice peak becomes similarly strong as the overdamped phonon in scattering intensity only below T* = 128.7 K.

IXS is not able to detect phonon intensities when there is strong elastic scattering, *e.g.*, at a reciprocal lattice point or at $\mathbf{q}_{CDW}$ (for T < T$_{CDW}$). Therefore, the CDW soft phonon mode in *2H*-TaSe$_2$ cannot be investigated anymore for $T \leq T_{CDW}$ due to the rise of the superlattice peak at $\mathbf{q}_{CDW}$. However, neighboring wave vectors are not affected by the superlattice peak. The phonon softening is detectable at $\mathbf{q} = (0.28,0,0)$ where the LA phonon energy softens by about 2 meV between T = 250 K and 130 K [Fig. 3(a)]. Cooling below *T*$_{CDW}$, we observe a hardening and narrowing of the phonon mode in the CDW phase. The temperature dependence of phonon energy levels off around T$_{C-IC}$ ≈ 90 K in reasonable agreement with results from Raman spectroscopy of the CDW soft phonon for T < T$_{CDW}$.[51]

In the following and Figure 6, we explain in detail the analysis of the momentum scans from which we deduce the correlation length of the CDW domains for $T_{CDW} < T < T^*$ [see Fig. 5(e)]. Momentum scans at zero energy transfer at $q$ = (0.3 – 0.37,0,0) were performed with the best momentum resolution possible on the HERIX spectrometer, $\Delta q$ = 0.018 Å$^{-1}$. Here, we decreased the effective size of the backscattering analyzers by closing a circular slit to a diameter of 18 mm (FWHM) compared to 95 mm opening in regular inelastic scans. From a fit we can determine the line width $\Gamma_{exp,FWHM}(T)$ [see Fig. 5(d)]. However, we cannot simply take the analysis of $\Gamma_{exp,FWHM}(T)$ at face value to analyze the CDW correlation length because our energy scans demonstrate the significant two-component nature of the scattering at zero energy transfer for $T > T_{CDW}$ [see Fig. 4(a)-(c)]. On the other hand, the energy scans [see Fig. 4] also show that the scattering from the soft phonon mode (1) dominates for $T \geq T^*$ and (2) is essentially constant for $T \leq T^*$. We conclude that scattering observed in momentum scans at zero energy transfer is mostly due to the soft phonon mode for $T \geq T^*$. Therefore, data taken at T = 129 K [blue circles in Fig. 6(a)] represent an estimate of the phononic background in our momentum scans and a fit [solid blue line in Fig. 6(a)] was subtracted from data taken at lower temperatures. The resulting scans indicate the evolution of the static CDW superlattice peak only [Fig. 6(b)]. Thereby, we can investigate the correlation length $\xi_{corr}$ of the static CDW superlattice peak at $T \leq T^*$ by approximating the resulting phonon-corrected momentum scans with a Voigt function. We fixed the Gaussian width of the Voigt function to the experimental resolution [see Fig. 5(d)]. The temperature dependences of the Lorentzian linewidth $\Gamma_{Lor,HWHM}(T)$ and the corresponding $\xi_{corr} = a/(2\pi \times \Gamma_{Lor,HWHM})$ are shown as black and red spheres in Figure 5(e), respectively. $\xi_{corr}$ increases below T* to about 200 Å just above $T_{CDW}$. Thus, the precursor phase in *2H*-TaSe$_2$ is characterized by medium-range-sized CDW domains, which only form a long-range CDW ordered state at $T \leq T_{CDW}$.

**Appendix B: Angle-resolved photoemission spectroscopy**

Angle-resolved photoemission spectroscopy (ARPES) is one of the most powerful experimental techniques to study the electronic band structure of solids. In CDW materials, ARPES has been indispensable to investigate the gap in the electronic excitation spectrum upon entering the ordered phase. Previous ARPES measurements in *2H*-TaSe$_2$ reported some anomalous behaviour: The band gap observed in energy distribution curves (EDCs) opens only below the onset of commensurate CDW order at $T_{C-IC} \approx$ 90 K.[26] Subsequent studies[27,52] corroborated this observation but reported the opening of a pseudo-gap already on cooling below $T_{CDW}$. The pseudo-gap opens in the bands of the K barrel and its size can be determined from the shift of the leading edge of the EDC. More details are given below in the analysis of our own data set.

In our experiment, we wanted to check the temperature evolution of the FS in *2H*-TaSe$_2$ in a narrow temperature range focusing on $T_{CDW}$ and T*. ARPES measurements at temperatures 132 K ≥ T ≥ 113 K were performed at the Bloch endstation of the R1 synchrotron at the MAX IV laboratory in Lund, Sweden, using linearly polarized light with 80 eV photon energy with a total resolution of < 10 meV in energy and < 0.2° in angle. The spot size was about 15 µm × 10 µm. The sample was cleaved using tape at 10$^{-8}$ mbar and measurements were performed at < 10$^{-10}$ mbar. Photoelectrons were recorded using a ScientaOmicron DA30-L hemispherical analyser. Sample orientation and cooling were achieved with a six axis "Carving" manipulator from SPECS GmbH cooled by a closed-loop liquid helium cryostat. Sample temperature was determined from previous calibration measurements. After each change in temperature setpoint, the sample temperature was allowed to stabilise for 20 minutes before the next measurement was performed. The sample used for ARPES measurements was a different single crystal from the same growth batch as that used for IXS measurements.

In agreement with previous studies[26,27], we observe hole-like circular FS sections centered on the $\bar{\Gamma}$ and $\bar{K}$ points and electron-like "dogbones" around the $\bar{M}$ point [Fig. 7(a)]. Here, $\bar{\Gamma}$, $\bar{K}$, and $\bar{M}$ denote the positions of high-symmetry points projected to the basal plane, i.e. $k_z$ = 0.

EDCs were taken from ARPES spectra centred around *M* along the $\bar{K}$-$\bar{M}$ direction [see blue dots #1 and #2 in Fig. 7(a)] as determined from Fermi surface maps taken at every temperature step. Each EDC is integrated over a 1° range. Each EDC was approximated with a single Gaussian peak modified by a Fermi-Dirac distribution with the Fermi level ($E_F$) kept constant across all temperatures in order to determine the $\bar{K}$ barrel peak positionand and the position of the leading edge for $\bar{K}$ barrel and $\bar{M}$ dogbone [***k*** corresponding to blue dot #1 in Fig. 7(a)]. We observe good agreement with the data down to at least -0.1 eV binding energy for all spectra. The pseudo-gap size is defined as the difference of the energies at half-height in EDCs taken on the $\bar{K}$ barrel [blue dot #1 in Fig. 7(a)] and the $\bar{M}$ dogbone [blue dot #2 in Fig. 7(a)]. Examples for T = 132 K and 113 K [Figs. 7(b) and (c)] are normalised for ease of comparison and the horizontal arrows denote the position at which the pseudo-gap size was defined. The obtained temperature dependences are summarized in Fig. 7(d). The shown values of the $\bar{K}$ barrel peak position (red dots, right-hand scale) and pseudogap size (black open squares, left hand scale) represent the average of the values obtained for the two pairs of $\bar{K}$ and $\bar{M}$ bands at positive and negative momentum relative to the $\bar{M}$ point visible in Fig. 7(a). The individual values show qualitatively the same behaviour across the studied temperature range as their average shown here. For both the $\bar{K}$ band peak position and the pseudo-gap size, we observe different slopes in their temperature dependence at low and high temperatures in our data set. Linear fits in each region [solid/dashed lines in Fig. 7(d)] cross close to $T_{CDW}$ = 121.3 K, i.e., the transition temperature deduced from x-ray momentum scans at zero energy transfer [see Fig. 4]. The observed kinks in the temperature dependence of the pseudogap and $\bar{K}$-band peak at $T_{CDW}$ agree also with previous results[27]. Finally, our analysis of the ARPES measurements reveals no particular electronic changes at T*.

52. Inosov, D. S. *et al.* Fermi surface nesting in several transition metal dichalcogenides. *New J Phys* **10**, 125027 (2008).

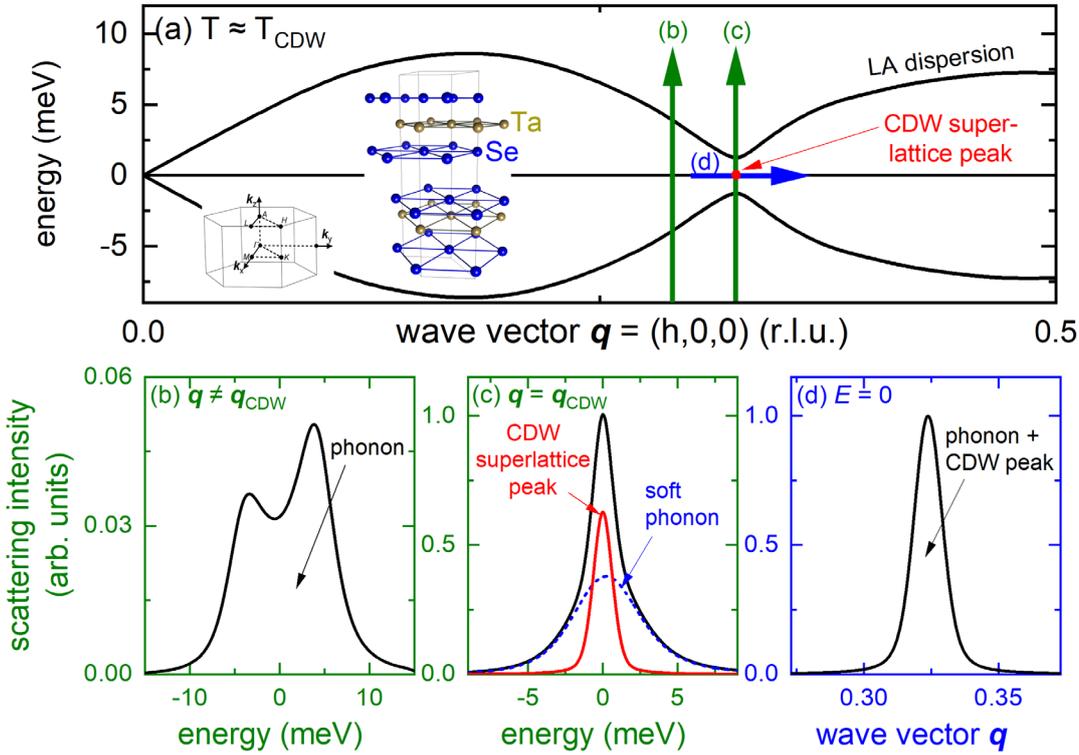

**FIG. 1. Charge-density-wave order in *2H*-TaSe$_2$**

**(a)** Schematic dispersion (black solid lines) of an acoustic phonon with a soft mode at $q_{CDW} \approx (1/3,0,0)$ for energy-loss (positive energies) and energy-gain (negative energies). The red dot indicates the position of the corresponding superlattice peak in the ordered phase. Thick vertical (green) and horizontal (blue) arrows illustrate the scans done on the HERIX spectrometer to investigate the phonon softening and superlattice peak formation in *2H*-TaSe$_2$. Labels (b),(c) and (d) refer to following panels showing typical results for the corresponding scans in more detail. Insets in (a) show the crystal structure of *2H*-TaSe$_2$ ($P6_3/mmc$, $a = b = 3.44$ Å, $c = 12.7$ Å, #194) and the layout of the Brillouin zone with high symmetry points labelled. **(b)** Energy scan at $q \neq q_{CDW}$. For a realistic picture we convoluted a damped harmonic oscillator function with the pseudo-voigt like experimental resolution ($\Delta E_{FWHM}$ = 1.5 meV). **(c)** Energy scan at $q = q_{CDW}$. The signal from the superlattice peak is approximated by the resolution function whereas the damped phonon is represented by a damped harmonic oscillator function convoluted with the resolution function. The scattering contributions from the superlattice peak [thick (red) solid line] are easily distinguished from the phonon contribution [thick (blue) dashed line]. **(d)** Momentum scan at zero energy transfer, E = 0, across the CDW superlattice peak [red dot in (a)]. Because of the finite energy resolution and the broad phonon lineshapes, it is not clear how strongly soft phonon mode and superlattice peak contribute to the scattering at zero energy transfer.

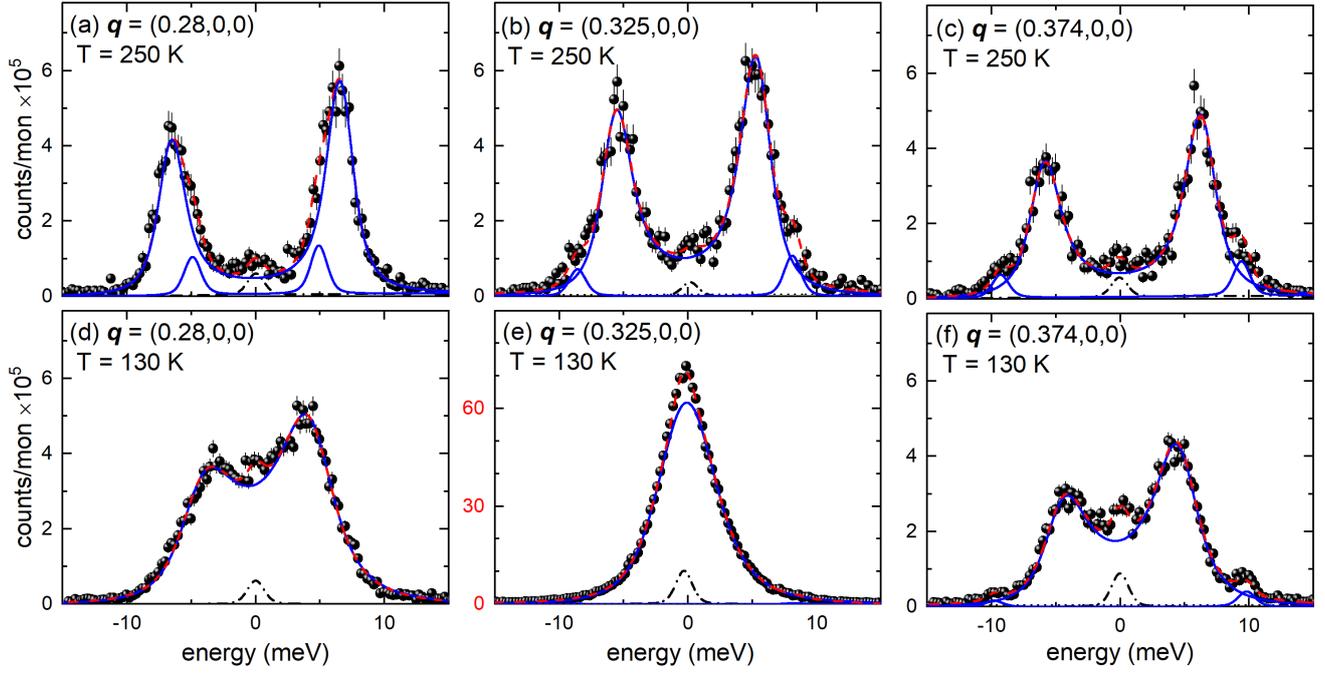

**FIG. 2. Soft phonon mode in *2H*-TaSe$_2$**
Energy scans taken at *q* ∥ [100] at **(a)-(c)** *T* = 250 K and **(d)-(f)** 130 K. Dashed (red) lines are fits consisting of DHO functions convoluted with the experimental resolution (blue solid lines), estimated background (straight dotted line) and a resolution limited pseudo-Voigt function for the elastic line (black dash-dotted line). Error bars represent s.d.

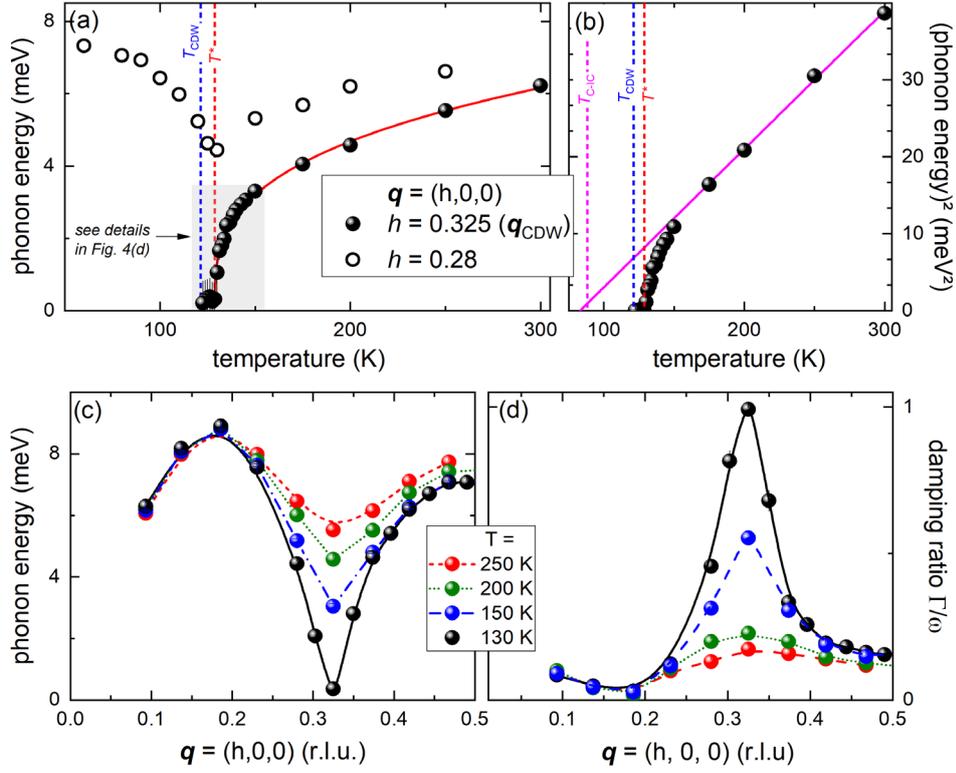

**FIG. 3. Soft phonon mode in *2H*-TaSe$_2$**
**(a)** Temperature dependence of the energies of the LA phonon modes at *q* = (0.325,0,0) (= *q*$_{CDW}$, spheres) and *q* = (0.28,0,0) (open circles). The line is a power law fit to the data for *q* = *q*$_{CDW}$ and T ≥ 130 K of the form |T-T*|$^δ$ yielding T* = 128.7 K (vertical red dashed line) and δ = 0.32 ± 0.02 [data in grey-shaded box are shown in more detail in Fig. 4(d)]. The vertical blue dashed line indicates T$_{CDW}$ = 121.3 K. **(b)** Square of the phonon energy at *q* = *q*$_{CDW}$ as function of temperature. The solid line is a linear fit for T ≥ 175 K. T$_{IC-C}$ ≈ 88 K indicates the transition temperature into the commensurate CDW phase [see inset in Fig. 5(c)]. **(c)** Dispersion and **(d)** damping Γ/ω of the LA mode along *q* = (h,0,0) for 250 K ≥ T ≥ 130 K. Lines are guides to the eye.

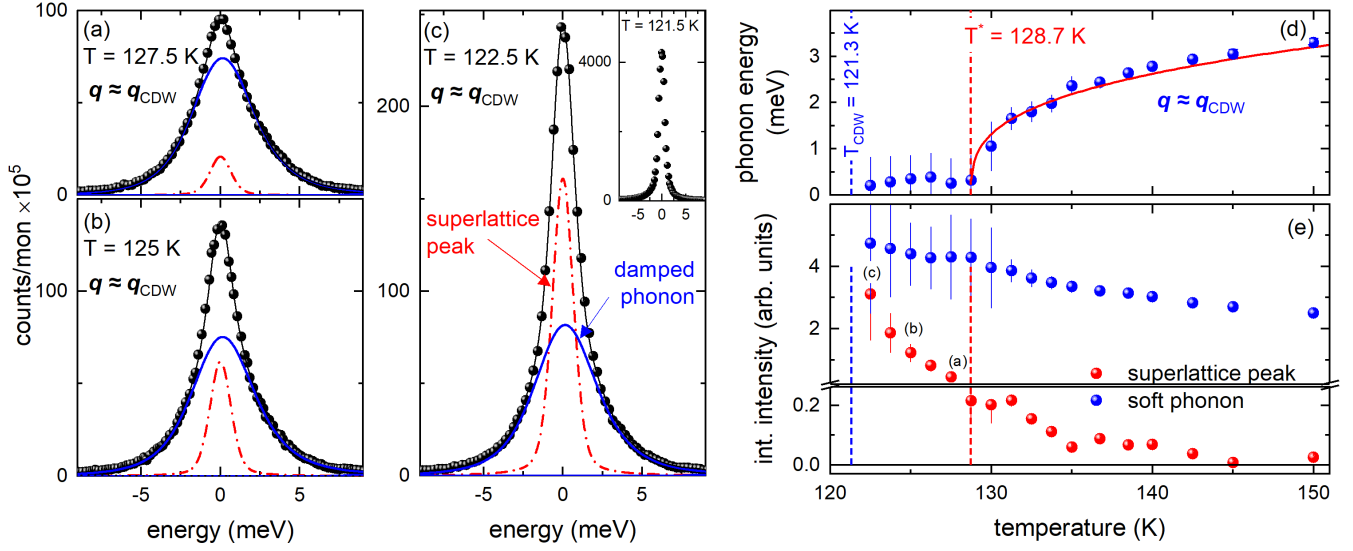

**FIG. 4. Evolution of elastic and inelastic scattering at $T_{CDW} \leq T \leq 150$ K**
**(a)-(c)** Energy scans taken at $Q = \tau + q$ with $\tau = (3,0,1)$ and $q = q_{CDW}$ [=(0.325,0,0)] at various temperatures 122.5 K $\leq T \leq$ 127.5 K, i.e, $T_{CDW} < T < T^*$. Thin solid (black) lines are fits consisting of DHO functions convoluted with the experimental resolution (blue solid lines), constant background and a resolution limited pseudo-Voigt function for the rising CDW superlattice peak (red dash-dotted line). The inset in (c) shows data taken at T = 121.5 K ($\approx T_{CDW}$). Here, no analysis of inelastic scattering is possible because of the increase of the CDW superlattice peak by more than a factor of 20. **(d)** Temperature dependent energy of the CDW soft phonon mode (blue spheres). The red solid line represents the same power law fit as shown in Fig. 2(c). **(e)** Temperature dependent integrated intensities of the CDW superlattice peak [red spheres, see also red dash-dotted lines in (a)-(c)] and the soft phonon mode [blue spheres, see also blue solid lines in (a)-(c)]. Small letters, (a)(b)(c), reference panels showing the corresponding IXS raw data. Vertical blue and red dashed lines [in (d)(e)] denote $T_{CDW} = 121.3$ K and $T^* = 128.7$ K, respectively.

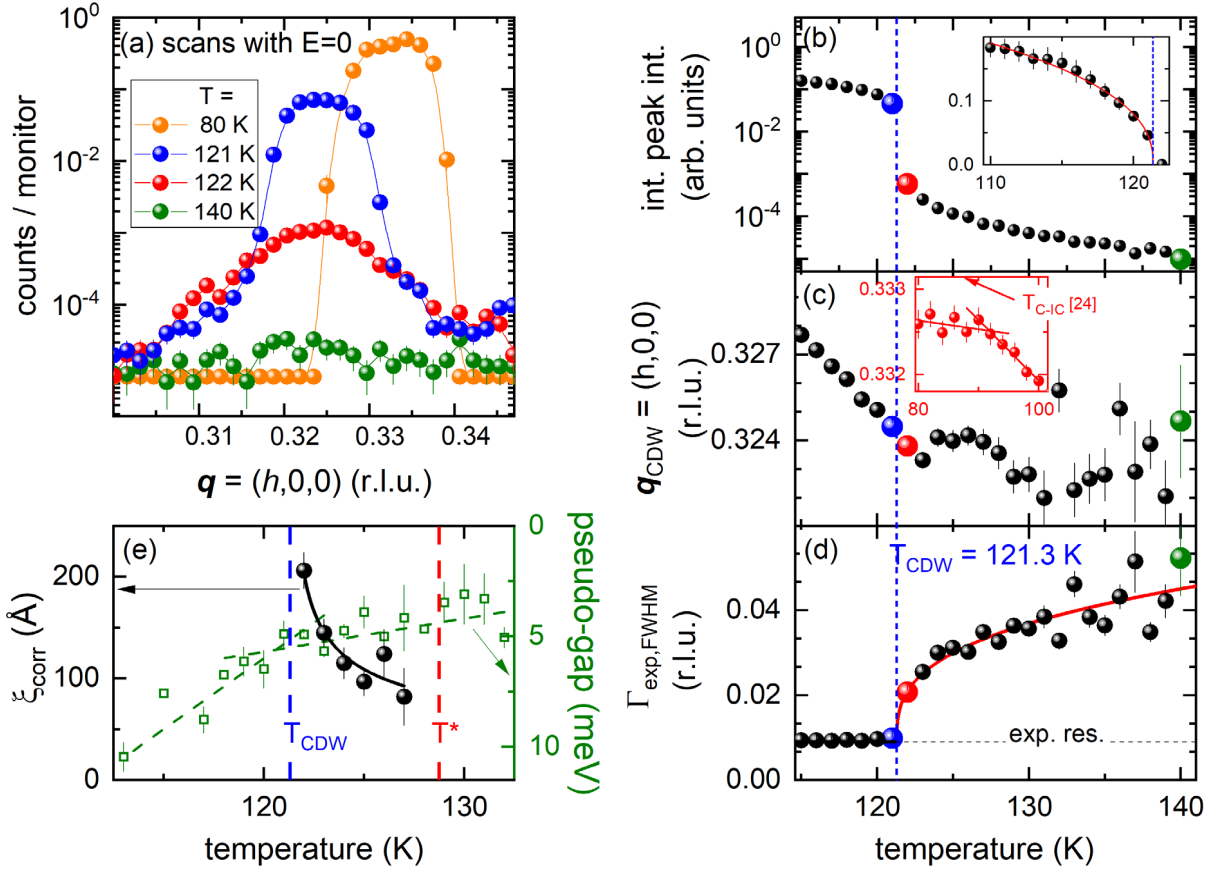

**FIG. 5. High-resolution momentum scans at E = 0 and temperatures T ≤ 140K.**
**(a)** Momentum scans at zero energy transfer along the [100] direction, i.e., across $q_{CDW}$ for temperatures above and below the CDW transition. Data were taken in a high-momentum resolution setup (see SM) to resolve the evolution of $q_{CDW}$ on cooling below $T_{CDW}$. **(b)-(d)** Temperature-dependent (b) integrated intensity, (c) position $q_{CDW}$ and (d) line width $\Gamma_{exp,FWHM}$ of peak fits to the momentum scans at zero energy transfer. Color-coded dots denote temperatures with data shown in (a). The inset in (b) shows the integrated intensity on a linear scale along with a power law fit (red line) of the form $|T-T_{CDW}|^\delta$ yielding $T_{CDW}$ = 121.3(2) K. The inset in (c) focuses on temperatures around $T_{C-IC} \approx 88$ K [29]. Lines are linear fits to the range below and above $T_{C-IC}$. The solid red line in (d) is another power law fit to the corresponding data for $T \geq 122$ K and yields $T_{CDW}$ = 121.3(2) K, indicated by the vertical blue dashed line. The dashed horizontal line in (d) denotes the experimental momentum resolution. **(e)** The temperature-dependent correlation length $\xi_{corr}$ (spheres) of the static scattering was obtained by subtracting the data at T = 129 K (≈ T*) as background and analyzing the linewidth of the static scattering rising below T* (see text, SM and Fig. S4). The solid line is a guide to the eye. The vertical blue and red dashed lines denote $T_{CDW}$ = 121.3 K and T* = 128.7 K, respectively. Green squares denote the temperature-dependent pseudo-gap deduced from ARPES measurements (see text, SM and Fig. 7). Dashed lines are linear fits to the data for $T < T_{CDW}$ and $T > T_{CDW}$.

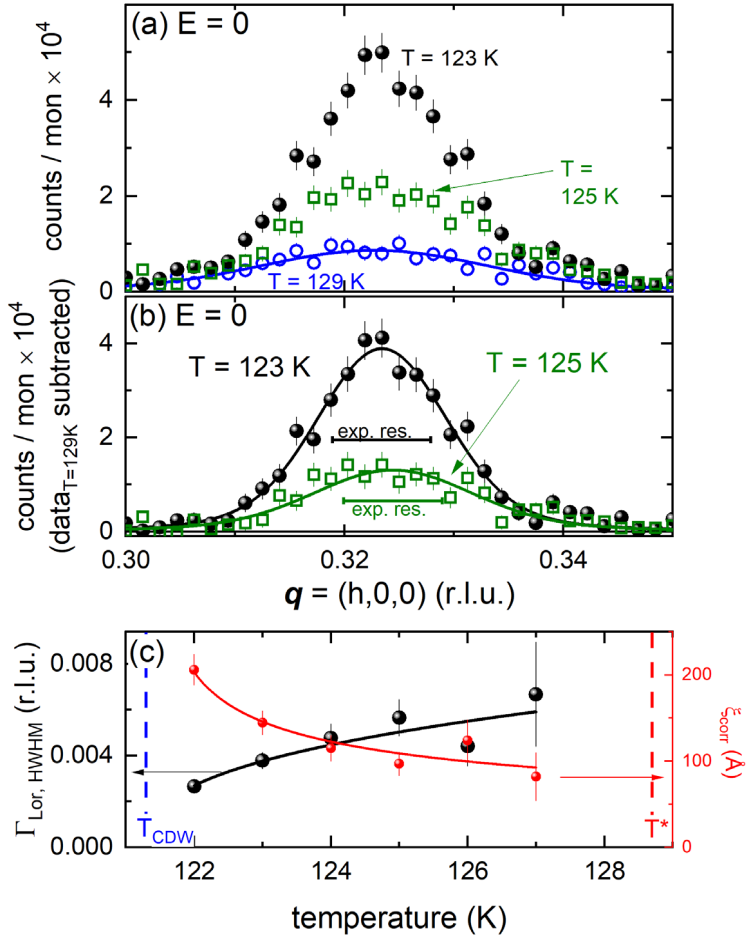

**Fig. 6**. (a) High resolution momentum scans at zero energy transfer for T = 123 K (black spheres) and 125 K (green open squares) and T =129 K (≈ T*, blue circles). The solid blue line denotes a fit to the data at T = 129 K and was subtracted from the data taken at lower temperatures as phonon dominated background (see text). (b) Background subtracted high resolution momentum scans at zero energy transfer for T = 123 K (black spheres) and 125 K (green open squares). Solid lines are approximated Voigt functions where the Gaussian widths were fixed to the experimental resolution (FWHM indicated by the horizontal bars). (c) Temperature-dependent line width of the Lorentzian part of the Voigt function, $\Gamma_{Lor,HWHM}$ (black spheres, left-hand scale). The corresponding correlation length $\xi_{corr} = a/(2\pi \times \Gamma_{Lor,HWHM})$ is shown in red (right-hand scale). Solid lines are guides to the eye. The vertical blue and red dashed lines denote $T_{CDW}$ = 121.3 K and T* = 128.7 K, respectively.

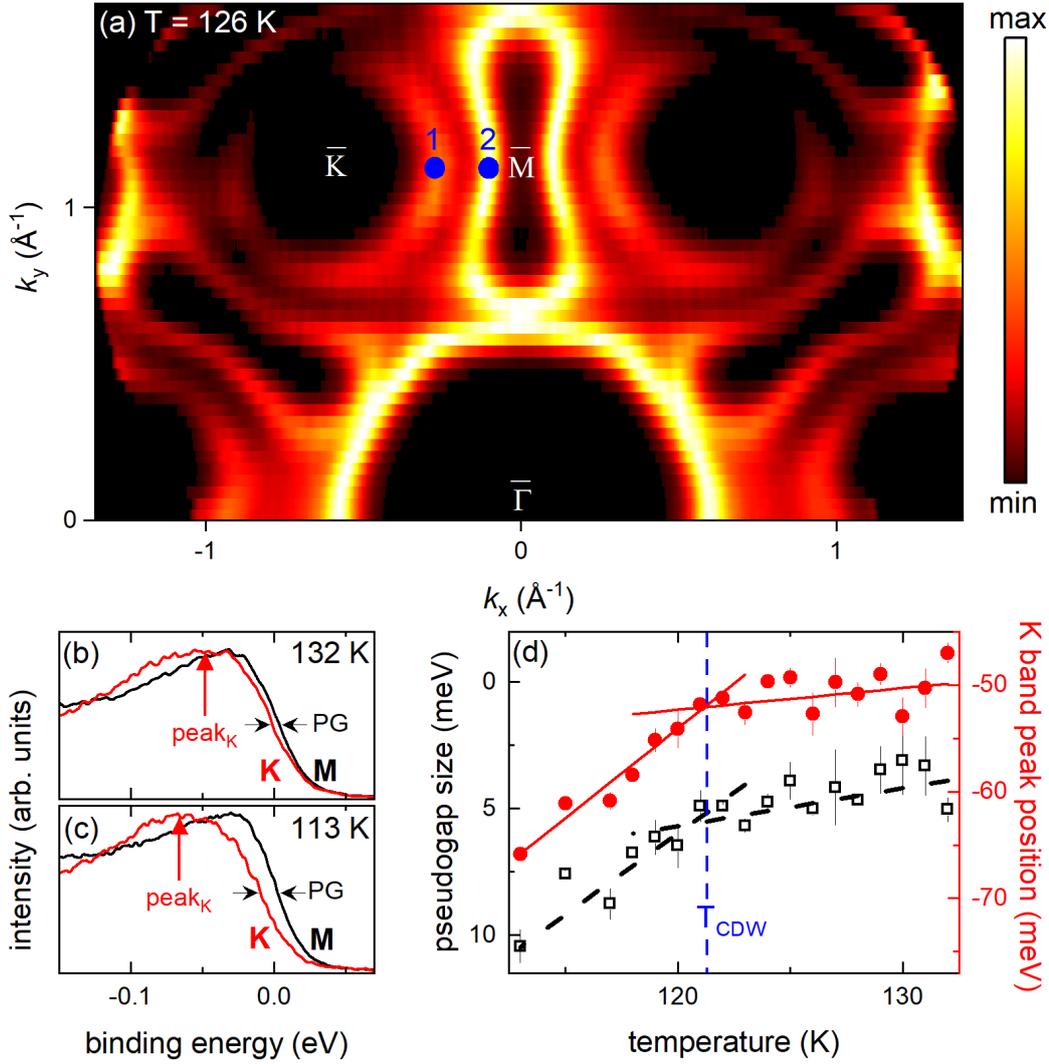

FIG. 7. (a) Fermi surface map measured at $T$ = 126 K. White characters denote high-symmetry points of the Brillouin zone. Blue points 1 and 2 indicate the momentum position of the EDCs shown in red and black, respectively, in panels (b) and (c). (b)(c) EDCs at (b) $T$ = 132 K and (c) 113 K obtained at the FS pockets around the $\bar{K}$ point [red solid line, blue point #1 in (a)] and the $\bar{M}$ point [black solid line, blue point #2 in (a)]. The horizontal and vertical arrows indicate the deduced sizes of the pseudo-gap and $\bar{K}$-band peak position, respectively. (d) Temperature dependence of the pseudo-gap (open squares) and $\bar{K}$-band peak position (dots). Lines are linear fits to the data for $T < T_{CDW}$ and $T > T_{CDW}$. The vertical blue dashed line denotes $T_{CDW}$ = 121.3 K deduced from elastic x-ray scattering.